\documentstyle[12pt]{article}
\newcommand{\nc}{\newcommand}
\nc{\rnc}{\renewcommand}
\rnc{\title}[1]{{\Large\bf\mbox{}\\\medskip#1\bigskip\medskip\\}}
\rnc{\author}[1]{{\large #1\smallskip\\}}
\nc{\address}[1]{{\em #1\medskip\\}}
\nc{\e}[1]{{\em #1\/}}
\nc{\comment}[1]{}
\nc{\itm}[2]{\\\noindent$\bullet$\ \ \e{#1}. \ #2}
\nc{\ru}[1]{\rule[-#1ex]{0ex}{#1ex}}
\rnc{\baselinestretch}{1.25}
\rnc{\arraystretch}{0.8}
\nc{\be}{\begin{equation}}
\nc{\ee}{\end{equation}}
\nc{\disp}{\displaystyle}
\nc{\ade}{\mbox{$A$-$D$-$E$}}
\nc{\calA}{{\cal A}}
\nc{\calH}{{\cal H}}
\nc{\calB}{{\cal B}}
\nc{\calC}{{\cal C}}
\nc{\calE}{{\cal E}}
\nc{\calI}{{\cal I}}
\nc{\calM}{{\cal M}}
\nc{\calN}{{\cal N}}
\nc{\calS}{{\cal S}}
\nc{\calV}{{\cal V}}
\def\jbar{{\overline{\jmath}}}
\def\qt{\tilde{q}}
\def\slhat{\widehat{s\ell}}

\def\rrangle{\rangle\!\rangle}

\def\slh{\widehat{sl}}

\def\({\left(}
\def\){\right)}
\font\tenmsb=msbm10 scaled \magstep1
\font\sevenmsb=msbm7 scaled \magstep1
\font\fivemsb=msbm5 scaled \magstep1
\newfam\msbfam
\textfont\msbfam=\tenmsb
\scriptfont\msbfam=\sevenmsb
\scriptscriptfont\msbfam=\fivemsb
\def\Bbb#1{{\fam\msbfam\relax#1}}

\begin{document}
\begin{titlepage}
\vspace*{\fill}
\rnc{\baselinestretch}{1.1}
\begin{center}
\title{On the Classification of Bulk and Boundary\\
Conformal Field Theories}
\medskip
\author{Roger E. Behrend}
\address{Institute for Theoretical Physics\\
State University of New York\\
Stony Brook, NY 11794-3840, USA}
\medskip
\author{Paul A. Pearce}
\address{Department of Mathematics and Statistics\\
University of Melbourne\\Parkville, Victoria 3052, Australia}
\medskip
\author{Valentina B. Petkova}
\address{Institute for Nuclear Research and Nuclear Energy\\
Tzarigradsko Chaussee 72\\ 1784 Sofia, Bulgaria}
\medskip
\author{Jean-Bernard Zuber}
\address{Service de Physique Th\'eorique, CEA-Saclay\\
 91191 Gif-sur-Yvette Cedex, France}
\bigskip\medskip

\begin{abstract}
\noindent  The classification of rational conformal field theories is
reconsidered from the standpoint of boundary conditions. Solving Cardy's
equation expressing the consistency condition on a cylinder is equivalent
to finding integer valued representations of the fusion algebra. A complete
solution not only yields the admissible boundary conditions but also gives
valuable information on the bulk properties. 
\end{abstract}
\end{center}
\vspace*{20mm}
\vspace*{\fill}
\end{titlepage}

The classification of conformal field theories (CFTs) remains an important
issue, both in the study of bulk and boundary critical phenomena~\cite{A98}
and in string theory~\cite{FS97}. The guiding principle is that of
consistency of the theory on an arbitrary $2D$ surface, with or without
boundaries.

To define a rational conformal field theory, one first specifies a chiral
algebra $\calA$, {\it e.g.} the Virasoro algebra or one of its extensions,
 at a
certain level.  Rationality means that at this level, $\calA$ has only a
finite set $\calI$  of admissible irreducible representations $\calV_i$,
$i\in \calI$. We denote by 
$\calV_{i^*}$ the representation conjugate to $\calV_i$ and $i=1$
refers to the vacuum representation. We suppose that the
characters $\chi_i(q)$ of these representations, the
symmetric  matrix ${S_i}^j$ of
modular transformations of the $\chi$'s and the fusion coefficients
${N_{ij}}^k$ of the $\calV$'s,  assumed to be given in terms of $S$ by the
Verlinde formula~\cite{Ve}, are all known.

A physical theory is then fully defined by the collection of bulk and
boundary fields and their 3-point couplings (``structure constants''). In
particular, the spectrum of bulk fields is described by the finite set Spec
of pairs $(j,\jbar)$ of representations  (with possible
multiplicities) of the left and right copies of
$\calA$, such that the Hilbert space of the theory on an infinitely long
cylinder reads $\calH=\oplus_{{\rm Spec}} \calV_j\otimes\calV_\jbar$. 
 We denote by $\calE$ the finite set of labels of the left-right
symmetric elements (up to conjugation) of the spectrum:
$\calE=\{j|(j,\jbar=j^*)\in\mbox{Spec}\}$, with these same multiplicities.
In terms
of all these data one is in principle able to compute exactly all
correlation functions of the CFT on an arbitrary $2D$ surface with or
without boundaries~\cite{CaL}.

These data, however, are subject to consistency constraints. Much emphasis
was originally put on bulk properties, namely on the consistency of the
4-point functions on the sphere~\cite{BPZ} and the zero-point function (the
modular invariant partition function) on a torus~\cite{Ca86}. In this
letter, we want to show that the consistency of the partition function on a
finite cylinder is equivalent to a well-posed algebraic problem. Once
solved, this not only determines the possible boundary conditions 
but also yields substantial information on the bulk properties, by
determining the diagonal part $\calE$ of the set Spec. The consistency 
condition on
a cylinder is the well known equation of Cardy~\cite{Ca89}, but it seems
that its consequences have never been fully exploited. 

We recall Cardy's discussion.  Let $W_n$ and $\overline{W}_n$ denote the 
spin $s_W$ generators of the $\calA$ chiral algebra acting on the left and 
right sectors. Then the
 $\calA$-invariant boundary states  satisfy the conditions
$(W_n-(-1)^{s_W} \overline{W}_{-n})|\phi\rangle=0$. (Here we assume
that the ``gluing automorphism''~\cite{RS,FS97} is trivial.) 
Solutions to this system of equations are spanned by special states   
$|j\rrangle$ (called Ishibashi states~\cite{I})  labelled
by the finite set $\calE$.  Let $Z_{AB}$ be the partition function of
the CFT on a cylinder of perimeter $T$ and length $L$ with boundary
conditions $A$ and
$B$. Regarded as resulting from the periodic ``time'' $T$ evolution of the
system with prescribed boundary conditions, 
it is a linear form in the characters with
integer coefficients: $Z_{AB}=\sum_{i\in \calI} {n_{iA}}^B \chi_i(q)$, 
$q=e^{-\pi T/L}$. 
It also results from the ``time'' $L$ evolution between states
$|A\rangle$ and $|B\rangle$: it is then a sesquilinear form in the
components of these boundary states on Ishibashi states. If we write 
\be 
|A\rangle=\sum_{j\in\calE} {{\psi_A}^j \over 
({{S_1}^j})^{1\over 2}}|j\rrangle \ , \label{Psi}  
\ee
then $Z_{AB}=\sum_{j\in \calE} 
{\psi_A}^j \({\psi_B}^j\)^*
\chi_j(\qt)/{S_1}^j$, $\qt=e^{-4\pi L/T}$, and Cardy's equation results 
from the identification of $\chi_i$ in these two alternative 
expressions of $Z_{AB}$:
\be
{n_{iA}}^B=\sum_{j\in \calE} 
{{S_i}^j\over {S_1}^j}\; {\psi_A}^j
\({\psi_B}^j\)^*\qquad\qquad {\rm for\ all\ }i\in\calI\ .\label{CardEqn}
\ee 
 
We have assumed that there is some quantity, 
for example a $Z_N$ charge, that discriminates representations with the
same character, {\it e.g.}  
 $\chi_i=\chi_{i^*}$,  and  enables one to write equation (\ref{CardEqn})
unambiguously. 
With the norm appropriate for boundary states~\cite{CaL,Ca89}, orthonormality
of $|A\rangle$ and $|B\rangle$ amounts to ${n_{1A}}^B=\sum_{j\in \calE}
 {\psi_A}^j
({\psi_B}^{j})^*=\delta_{AB}$. Reality of $Z_{AB}$ implies that
${n_{iA}}^B={n_{i^*B}}^A$. It is also natural to introduce conjugate states
$|A^*\rangle$ such that $ {\psi_{A^*}}^{j}={\psi_A}^{j^*}=({\psi_A}^{j})^*$.
Suppose we have found a complete set of boundary states,
i.e. such that $\sum_B {\psi_B}^j \({\psi_B}^{j'}\)^*=\delta_{jj'}$. 
Then, using
the fact that the ratios ${{S_i}^j\over {S_1}^j}$ for a given $j$ form a
representation of the fusion algebra, as a consequence of the Verlinde
formula: ${{S_{i_1}}^j\over {S_1}^j}{{S_{i_2}}^j\over {S_1}^j}=\sum_{i_3}
{N_{i_1 i_2}}^{i_3} {{S_{i_3}}^j\over {S_1}^j}$, it follows that the
matrices $n_i=({n_{iA}}^B)$ also form a representation of the fusion algebra
\be
\sum_B {n_{i_1A}}^B {n_{i_2B}}^C=
\sum_{i_3} {N_{i_1 i_2}}^{i_3}{n_{i_3A}}^C\ .\label{FuseEqn}
\ee
Equation (\ref{CardEqn}) expresses $n_i$ in terms of its eigenvalues
${{S_i}^j\over {S_1}^j}$ and its eigenvectors $\psi^j$. The matrices
commute and are normal ($n_i$ commutes with $n_i^T=n_{i^*}$).

Thus the problem of finding all the orthonormal and complete solutions to
Cardy's equation (\ref{CardEqn}) is equivalent to that of finding all
matrix representations of the fusion algebra with nonnegative integer
coefficients that satisfy $n_i^T=n_{i^*}$.
The fact that classes of 
boundary partition functions are associated with
representations of the fusion algebra has been recognized 
before~\cite{Ca89,DFZ12, PSS}. To the best of our knowledge, however, the
general and simple argument above has not been given. 

The search for  representations of the fusion algebra
is a well posed problem that may be approached by algebraic or
combinatorial methods. As the fusion algebra has a finite number of
generators, one first determines the representations of these
generators. The latter are matrices whose possible eigenvalues
${{S_i}^j\over {S_1}^j}$ are known. They may be regarded as the adjacency
matrices of graphs that characterize the representation.  For 
theories with the affine (current) algebra $\slhat(2)$ 
as a chiral algebra, this problem has been solved long ago~\cite{DFZ12}.
The representations of $\slhat(2)$ at level $k$ are labelled by an integer
$1\le j\le k+1$, and Cardy's equation says that the generator $n_2=n_{2^*}$ has
eigenvalues ${{S_2}^j\over {S_1}^j}=2 \cos {\pi j\over k+2}$. The only
symmetric indecomposable matrices with nonnegative integer entries and
eigenvalues less than $2$ are the adjacency matrices of $A$-$D$-$E$ Dynkin
diagrams and of the ``tadpole'' graphs $A_{2n}/{\Bbb Z}_2$~\cite{GHJ}. Only
the former solutions are retained as their spectrum matches the spectrum of
$\slhat(2)$ theories, known by their modular invariant partition
functions~\cite{CIZ}. For a theory classified by a Dynkin diagram $G$ of
$A$-$D$-$E$ type, the set $\calE$ is the set of Coxeter exponents of $G$. 
The matrices $n_i$ are then the ``fused adjacency matrices'' 
or ``intertwiners''
defined recursively by equation (\ref{FuseEqn}): $n_{i+1}=n_2 n_i-n_{i-1}$,
$i=2,3,\ldots,k$; $n_1=I$~\cite{DFZ12,PZh}, and one verifies that all their
entries are nonnegative integers.  This set of complete
orthonormal solutions of Cardy's equation for $\slhat(2)$ theories
is unique up to a relabelling of the states $|A\rangle$. Particular
solutions in the $D_{{\rm odd}}$ cases were obtained by a different
method in~\cite{PSS}.

For minimal $c<1$ theories, if $c=1-{6(g-h)^2\over gh}$ and the theory is
classified by a pair $(A_{h-1},G)$, $h$ odd, $G$ a Dynkin 
diagram of $A$-$D$-$E$
type with Coxeter number $g$, a class of solutions is obtained by tensor
products of the solutions of the $\slhat(2)$ case: $n_{(rs)}=N_r\otimes
n_s$ where $r=1,3,\ldots,h-2$, $s=1,\ldots,g-1$, 
$N_r$ are the fusion matrices
of $\slhat(2)$ at level $h-2$ and $n_s$ are the intertwiners just
mentioned pertaining to $G$~\cite{BePZ}. It is likely that all (orthonormal
and complete) solutions of Cardy's equation are of this type.  We have 
checked it explicitly in the cases of $G=D_4, D_6, E_7$. 
The issue of completeness of boundary conditions in the three-state
Potts model and other minimal models
has also been addressed recently in~\cite{AOSFS98}.

For more complicated CFTs, like those based on higher rank affine algebras,
no general result on the representations of the fusion algebra is known,
although a few steps have been taken~\cite{DFZ12,PZ12}. See also 
some recent more abstract work in this direction~\cite{XuBE}. 

The considerations of the present paper  put on a firmer
ground  the general program of classification of CFTs 
through the classification of ${\Bbb N}$-valued matrix 
representations of the fusion
algebra proposed in~\cite{DFZ12}.  
For a given chiral algebra $\calA$ and an ${\Bbb N}$-valued matrix
representation $\{n_i\}$ of its fusion algebra, the commuting and
normal matrices $n_i$ may be diagonalized in an orthonormal basis
${\psi}^j$ labelled by representations $\calV_j$ of $\calA$, with
possible multiplicities, which according to Cardy's equation must give the
diagonal part $\calE$ 
of the spectrum of the theory
in the bulk. It then must be a relatively easy task to decide if this
diagonal subset may be supplemented to make a fully consistent theory in
the bulk, namely if
\be
\sum_{j\in \calE} |\chi_j(q)|^2 +\mbox{off-diagonal terms}
\ee
may be made modular invariant. 
For example, in $\slhat(2)$ theories, as discussed above, from the possible
solutions to Cardy's equation, one recovers the classification of modular
invariants, once solutions of type $A_{2n}/{\Bbb Z}_2$ have been discarded.
For higher rank, $\slhat(3)$ for example, it is also known that some
representations of the fusion algebra do not give rise to a modular
invariant~\cite{DFZ12}.  In some cases~\cite{DFZ12,PZ12}, there
are explicit expressions of the modular invariant partition function 
in terms of the matrices $n_i$. 
In block diagonal cases, $Z=\sum_{B\in T} |\sum_i n_{i1}{}^B \chi_i|^2$, 
for a special subset of boundary conditions 
$T$ and a special boundary state denoted $1\in T$. 

The eigenvectors $\psi$ carry also physical information on boundary 
and bulk properties. The $g$-factors introduced by Affleck 
and Ludwig~\cite{AfL}
giving the groundstate degeneracies
are easily seen to be (in a unitary theory) $g_A=\psi_A^1/\sqrt{{S_1}^1}$.
The bulk--boundary 
reflection coefficients~\cite{CaL} are expressed in our notation as 
\be
\sqrt{{S_1}^j\over {S_1}^1}\  C_{(j,j^*), 1}^A={{\psi_A}^j\over{\psi_A}^1}
\qquad\qquad j\in \calE \ .
\ee
These ratios provide the $1$-dimensional representations of
 the Pasquier algebra~\cite{Pa} with structure constants
\be
{M_{ij}}^k= \sum_A{{\psi_A}^i{\psi_A}^j({\psi_A}^{k})^*\over{\psi_A}^1}\ .
\ee
In $\slh(2)$ and minimal models, these ${M_{ij}}^k$ are,  
in a suitable normalization, the relative OPE coefficients in the bulk 
${C_{(ii^*)(jj^*)}}^{(kk^*)}$, as seen by the independent study 
of bulk~\cite{PZ12} and boundary~\cite{CaL} locality equations.
In the diagonal cases $\calE =\calI$, we have ${M_{ij}}^k = {N_{ij}}^k$.

It is thus suggested that the
classification of ${\Bbb N}$-valued representations of the fusion algebra
is a profitable route to the classification of CFTs and determination of
a large part of their data.

\section*{Acknowledgements}
REB and PAP are supported by the Australian Research Council. VBP
acknowledges the support and hospitality of ICTP, Trieste, and the 
partial support of the Bulgarian National Research Foundation 
 (contract $\Phi$-643). J-BZ
acknowledges partial support of the EU Training and Mobility of Researchers
Program (Contract FMRX-CT96-0012), which made possible an extremely
profitable stay in SISSA, Trieste.
VBP and J-BZ would like to thank J. Fuchs, V. Schomerus and C. Schweigert
for inviting them to a very enjoyable meeting on ``Conformal field theory
of D-branes'' at DESY,  where Ya.\ Stanev presented a different approach to 
equation (\ref{CardEqn}), and where they also had stimulating discussions 
with them as well as with A. Recknagel, A. Sagnotti and Ya.\ Stanev.


\newpage

\begin{thebibliography}{99}\label{bib}

\bibitem{A98} Affleck, I., talk at StatPhys20, Paris, July 1998.

\bibitem{FS97} Fuchs, J. and  Schweigert, C., {\em Branes: From 
Free Fields
to General Backgrounds}, hep-th 9712257, and further references therein.

\bibitem{Ve} Verlinde, E., Nucl.\ Phys.\ {\bf B300} (1988) 360.

\bibitem{CaL} 
Cardy, J. L. and Llewellen, D. C., Phys.\ Lett.\ {\bf B259} (1991) 274; 
Llewellen, D.~C., Nucl.\ Phys.\ {\bf B372} (1992) 654.

\bibitem{BPZ} Belavin, A. A., Polyakov, A. M. and Zamolodchikov, A. B.,
Nucl.\ Phys.\ {\bf B241} (1984) 333.

\bibitem{Ca86} Cardy, J. L., Nucl.\ Phys.\ {\bf B270} (1986) 186.

\bibitem{Ca89} Cardy, J. L., Nucl.\ Phys.\ {\bf B324} (1989) 581.

\bibitem{RS} Recknagel, A. and Schomerus, V., {\em D-branes in Gepner
models}, hep-th 9712186.

\bibitem{I} Ishibashi, N., Mod.\ Phys.\ Lett.\ {\bf A4} (1989) 251.

\bibitem{DFZ12} Di Francesco, P. and  
Zuber, J.-B., Nucl.\ Phys.\ {\bf B338} (1990) 602;  
in  ``Recent Developments in Conformal Field Theories'', 
S. Randjbar-Daemi et al.\ eds., World Scientific  1990.

\bibitem{PSS} Pradisi, G., Sagnotti, A. and Stanev, Ya.\ S.,  
Phys.\ Lett.\ {\bf B381} (1996) 97; Sagnotti, A. and Stanev, Ya. S., 
{\it Open descendants in conformal field theory}, hep-th 9605042.

\bibitem{GHJ} Goodman, F. M., de la Harpe, P. and Jones, V. F. R., {\em
Coxeter Dynkin diagrams and towers of algebras}, Vol.\ 14, MSRI Publications,
Springer, Berlin 1989.

\bibitem{CIZ} Cappelli, A., Itzykson, C. and  Zuber, J.-B.,
Comm.\ Math.\ Phys.\ {\bf 113} (1987) 1.

\bibitem{PZh} Pearce, P. A. and  Zhou, Y.-K., 
Int.\ J. Mod.\ Phys.\ {\bf B7} (1993) 3649;  {\bf B8} (1994) 3531.

\bibitem{BePZ} Behrend, R. E., Pearce, P. A. and Zuber, J.-B., {\em
Integrable boundaries, conformal boundary conditions and $A$-$D$-$E$ fusion
rules}, hep-th 9807142.
 
\bibitem{AOSFS98}  Affleck, I., Oshikawa, M. and Saleur, H., 
{\em Boundary critical 
phenomena in the three-state Potts model}, cond-mat 9804117;
Fuchs, J. and Schweigert, C., {\em Completeness of boundary 
conditions for the critical three-state Potts model}, hep-th 9806121.
       
 \bibitem{PZ12} Petkova, V. B. and Zuber, J.-B., Nucl.\ Phys.\ {\bf B438} 
(1995) 347; Nucl.\ Phys.\ {\bf B463} (1996) 161;  
{\em Conformal Field Theory and Graphs}, hep-th 9701103.

\bibitem{XuBE}  Xu,  F., 
Comm.\ Math.\ Phys.\ {\bf 192} (1998) 349;
B\"ockenhauer J. and Evans,~D.~E., {\it Modular invariants, graphs
and $\alpha$-induction for nets of subfactors}, hep-th 9805023.

\bibitem{AfL} Affleck, I. and
Ludwig, A. W. W., Phys.\ Rev.\ Lett.\ {\bf 67} (1991) 161. 

\bibitem{Pa} Pasquier, V., J. Phys.\ {\bf A20} (1987) 5707.
\end{thebibliography}
\end{document}